\documentclass{rspublic}
\usepackage{graphicx}
\usepackage{mathbbol}
\usepackage{verbatim}
\usepackage{natbib}
\def\Tr{\mbox{Tr}}

\begin{document}

\title[Quantum Fluctuation Theorems]{
Quantum Bochkov-Kuzovlev Work Fluctuation Theorems}
\author[M. Campisi, P. Talkner and P. H\"anggi]{Michele Campisi, Peter
  Talkner and Peter H\"anggi}
\affiliation{Institut f\"ur Physik,
  Universit\"at Augsburg,
  Universit\"atsstr. 1,
  D-86135 Augsburg, Germany}

\label{firstpage}

\maketitle

\begin{abstract}{Fluctuation relations, nonequilibrium thermodynamics,
    response theory, quantum work, canonical, microcanonical, entropy}
The quantum version of the Bochkov-Kuzovlev identity is derived on
the basis of the appropriate definition of work as the difference of
the measured internal energies of a quantum system at the beginning
and at the end of an external action on the system given by a
prescribed protocol. According to the spirit of the original
Bochkov-Kuzovlev approach, we adopt the ``exclusive'' viewpoint,
meaning that the coupling to the external work-source is {\it not}
counted as part of the internal energy. The corresponding canonical
and microcanonical quantum fluctuation theorems are derived as well,
and are compared to the respective theorems obtained within the
``inclusive'' approach. The relations between the quantum
inclusive-work $w$, the exclusive-work $w_0$ and the dissipated-work
$w_{dis}$, are discussed and clarified. We show by an explicit
example that $w_0$ and $w_{dis}$ are distinct stochastic quantities
obeying different statistics.
\end{abstract}

%%%%%%%%%%%%%%%%%%%%%%%%%%%%%%%%%%%%%%%%%%%%%%%%%%%%%%%%%%%%%%%%%%%%%%%%
\section{\label{sec:intro}Introduction}
%%%%%%%%%%%%%%%%%%%%%%%%%%%%%%%%%%%%%%%%%%%%%%%%%%%%%%%%%%%%%%%%%%%%%%%%
One of the main objectives of {\it nonequilibrium thermodynamics} is
the study of the response of physical systems to applied external
perturbations. Around the mid of the last century major advancements
were obtained in this field with the development of linear response
theory by several authors, among which we mention
\cite{CallenWelton51,Green_JChemPhys52,Kubo57}. This theory,
inspired by the works of \citet{EinsteinBook} on the Brownian
movement and of \citet{Johnson_PR28} and \citet{Nyquist_PR28} on
noise in electrical circuits, established that, under certain
circumstances, the linear response to small perturbations is
determined by the equilibrium fluctuations of the system. In
particular, the linear response coefficients are proportional to
two-point correlation functions for Hamiltonian systems
\citep{Kubo57} as well as for stochastic, generally non-equilibrium
systems \citep{HanggiThomas82}. In principle, an infinite hierarchy
of higher order fluctuation-dissipation relations connects the
$n$-th order response coefficients to $(n+1)$-point correlation
functions.

In contrast, fluctuation theorems are compact relations that provide
information about the fully {\it nonlinear response}. Accordingly,
fluctuation-dissipation relations of all orders can be derived
therefrom. \citet{BochkovKuzovlev77,BochkovKuzovlev81} were the
first to put forward one such fully nonlinear fluctuation theorem.
These authors noticed that, for a {\it classical} system, their
general fluctuation theorem implies the following, extremely simple
nonequilibrium identity:
\begin{equation}
 \langle e^{-\beta W_0}\rangle =1
\label{eq:BKE}
\end{equation}
where $W_0$ is the work done on the system by the external
perturbation during one specific realization thereof, $\langle \cdot
\rangle$ denotes the average over many realizations of the same
perturbation, and $\beta=(k_B T)^{-1}$, with $T$ the initial
temperature of the system and $k_B$ Boltzmann's constant. Due to the
properties of convexity of the exponential function, an almost
immediate consequence of (\ref{eq:BKE}), is the second law of
thermodynamics in the form, $\langle W_0 \rangle \geq 0$; i.e. when
a system is perturbed from an initial thermal equilibrium, on
average, it can only absorb energy.

The works of \citet{BochkovKuzovlev77,BochkovKuzovlev81} has
recently re-gained a great deal of attention, after \citet{Jarz97}
derived, within the framework of classical mechanics, a salient
result similar to Eq. (\ref{eq:BKE}):
\begin{equation}
 \langle e^{-\beta W}\rangle = e^{-\beta \Delta F}
\label{eq:JE}
\end{equation}
which, in contrast to Eq. (\ref{eq:BKE}), allows to extract an
equilibrium property of the system, i.e., its free energy
(difference) $F$, from nonequilibrium fluctuations of work $W$.
Evidently, the definitions of work adopted by \citet{Jarz97} and
\citet{BochkovKuzovlev77,BochkovKuzovlev81} (denoted here
respectively as $W$ and $W_0$) do not coincide. The relationships
between these two work definitions and the corresponding
nonequilibrium identities, Eqs. (\ref{eq:BKE},\ref{eq:JE}), were
discussed in a very clear and elucidating manner in
\citep{Jarz_CRPhys07,Jarz_JStatMech07}, which, for the sake of
clarity, we summarize below.

Let us express the time dependent
Hamiltonian of the driven classical system as the sum of the
unperturbed system Hamiltonian $H_0$ and the interaction energy
stemming from the coupling of the external time dependent
perturbation $X(t)$ to a certain system observable $Q$:
\begin{equation}
H(\mathbf{q},\mathbf{p};t)= H_0(\mathbf{q},\mathbf{p}) -
X(t)Q(\mathbf{q},\mathbf{p}),
\label{eq:H-cl}
\end{equation}
We restrict ourselves to the simplest case of a protocol governed by a
single ``field'' $X(t)$ coupling to the conjugated generalized
coordinate $Q$.
Generalization to several fields $X_i(t)$ coupling to different
generalized coordinates $Q_i$ is straightforward.

The definition of work $W$ according to \citet{Jarz97} stems from
an {\it inclusive} viewpoint, where one counts the term $X(t)Q$ as
being part of the system internal energy. In contrast the
definition of work $W_0$ according to
\citet{BochkovKuzovlev77,BochkovKuzovlev81}
belongs to an {\it exclusive} viewpoint where instead, such term
is not counted as part of the energy of the system. More explicitly,
if $\mathbf{q}_0,\mathbf{p}_0$ is a certain initial condition that
evolves to $\mathbf{q}_f,\mathbf{p}_f$ in a time $t_f-t_0$, according
to the Hamiltonian evolution generated by $H$, then the two different
definitions of work become:
\begin{align}
 W &\doteq  H(\mathbf{q}_f,\mathbf{p}_f;t_f)-
 H(\mathbf{q}_0,\mathbf{p}_0;t_0)\\
 W_0 & \doteq H_0(\mathbf{q}_f,\mathbf{p}_f)-
 H_0(\mathbf{q}_0,\mathbf{p}_0)
\end{align}

It is important to stress that \citet{BochkovKuzovlev77,BochkovKuzovlev81} only
obtained Eq. (\ref{eq:BKE}) in the {\it classical} case,
notwithstanding the fact that they developed a quantum version of
their theory, as well. This difficulty is related to the fact that
work was identified by \citet{BochkovKuzovlev77,BochkovKuzovlev81} with the
quantum
expectation of a pretended {\it work operator}, given by the
difference of final and initial Hamiltonian in the
Heisenberg representation.
To be more clear, if the quantum Hamiltonian reads:
\begin{equation}
  H(t)= H_0-X(t) Q
\label{eq:H}
\end{equation}
where now $ H, H_0$ and $ Q$ are hermitian operators, the work
operator was defined by \citet{BochkovKuzovlev77,BochkovKuzovlev81} as:
\footnote{\citet{BochkovKuzovlev77} defined the ``operator of energy absorbed by
the system'' $E=\int_{t_0}^{t_f}X(\tau)Q^H(\tau)d\tau$, where $Q^H(\tau)$ is the
operator $Q$ in the Heisenberg representation. It is not difficult to prove that
$E$ coincides with $W_0$ in Eq. (\ref{eq:work-op-ex}).}
\begin{equation}
  W_0 \doteq   H_0^H(t_f)- H_0
\label{eq:work-op-ex}
\end{equation}
where the superscript $H$ denotes Heisenberg picture.
A similar approach was employed also within the inclusive viewpoint,
with work defined as \citep{Allahverdyan_PRE05}
\begin{equation}
  W =  H^H(t_f)- H_0
\label{eq:work-op-in}
\end{equation}
As pointed out clearly by some of us before with the work in
\citep{Lutz_PRE07} the Jarzynski Equality (\ref{eq:JE}) {\it cannot}
be obtained on the basis of the work operator (\ref{eq:work-op-in}).
Likewise the Bochkov-Kuzovlev identity (\ref{eq:BKE}) cannot be
obtained on the basis of Eq. (\ref{eq:work-op-ex}). It is by now
clear that the impossibility of extending the classical results
(\ref{eq:BKE},\ref{eq:JE}) on the basis of quantum work operators
(\ref{eq:work-op-ex},\ref{eq:work-op-in}), respectively, is related
to the fact that work characterizes a process, rather than a state
of the system, and consequently cannot be represented by an
observable that would yield work as the result of a single
projective measurement. In contrast, energy must be measured twice
in order to determine work, once before the protocol starts and a
second time immediately after it has ended. The difference of the
outcomes of these two measurements then yields the work performed on
the system in a particular realization \citep{Lutz_PRE07}.

In this paper we will adopt the exclusive viewpoint of
\citet{BochkovKuzovlev77,BochkovKuzovlev81}, but use the proper
definition of work as the difference between the outcomes of two
projective measurements of $H_0$, to obtain the quantum version of
Eq. (\ref{eq:BKE}). Indeed we will develop the theory of quantum
work fluctuations within the exclusive two-point measurements
viewpoint in great generality. In Sec. \ref{sec:CFW} we study the
characteristic function of work.  In Sec. \ref{sec:can} and Sec.
\ref{sec:mu-can} we derive the exclusive versions of the
Tasaki-Crooks quantum fluctuation theorem
\citep{TH_JPA07,TCH_JStatMech09,CTH_PRL09}, and of the
microcanonical quantum fluctuation theorem \citep{THM_PRE08},
respectively. Sec. \ref{sec:discussion} closes the paper with some
remarks concerning the relationships between the inclusive-work, the
exclusive-work, and the dissipated-work.

%%%%%%%%%%%%%%%%%%%%%%%%%%%%%%%%%%%%%%%%%%%%%%%%%%%%%%%%%%%%%%%%%%%%%%%%
\section{\label{sec:CFW}Characteristic function of work}
%%%%%%%%%%%%%%%%%%%%%%%%%%%%%%%%%%%%%%%%%%%%%%%%%%%%%%%%%%%%%%%%%%%%%%%%
As mentioned in the introduction, work is properly defined in
quantum mechanics as the difference of the energies measured at
the beginning and end of the protocol, i.e., at times $t_0$ and
$t_f>t_0$, respectively.
According to the exclusive
viewpoint that we adopt here, this energy is given by the
unperturbed Hamiltonian $H_0$. Let $e_{n}$ and
$|n,\lambda\rangle$, denote the eigenvalues and eigenvectors of
$H_0$:
\begin{equation}
 H_0 |n,\lambda \rangle=e_{n} |n,\lambda\rangle\, .
\end{equation}
Here $n$ is the quantum number labeling the eigenvalues of $H_0$ and
$\lambda$ denotes further quantum numbers needed to specify uniquely
the state of the system, in case of degenerate energies. A
measurement of $H_0$ at time $t_0$ gives a certain eigenvalue $e_n$
while a subsequent measurement of $H_0$ at time $t_f$ gives another
eigenvalue $e_m$, so that
\begin{equation}
 w_0 = e_m-e_n
\end{equation}
Evidently $w_0$ is a stochastic variable due to the intrinsic
randomness entailed in the quantum measurement processes and
possibly in the statistical mixture nature of the initial
preparation. In the following we derive the statistical properties
of $w_0$, in terms of its probability density function (pdf), and the
associated characteristic function of work.

Let the system be prepared at time $t<t_0$ in a certain state
described by the density matrix $\rho(t_0)$. We further assume
that the perturbation $X(t)$ is switched on at time $t_0$. At the
same time the first measurement of $H_0$ is performed, with
outcome $e_n$. This occurs with probability:
\begin{equation}
 p_n= \sum_\lambda \langle n, \lambda| \rho(t_0)|n, \lambda \rangle=
 \Tr \, P_n \rho(t_0)
\end{equation}
where $P_n$ is the projector onto the eigenspace spanned by the
eigenvectors belonging to the eigenvalue $e_n$:
\begin{equation}
 P_n = \sum_{\lambda} |n,\lambda\rangle \langle n, \lambda |
\end{equation}
and $\Tr$ denotes trace over the Hilbert space.
According to the postulates of quantum mechanics, immediately
after the measurement the system is found in the state:
\begin{equation}
 \rho_n=P_n \rho(t_0) P_n /p_n \, .
\end{equation}
For times $t>t_0$ the system evolves according to
\begin{equation}
\rho(t)=U_{t,t_0}\rho_n U^{\dagger}_{t,t_0}
\end{equation}
with $U_{t,t_f}$ denoting the unitary time evolution operator obeying the
Schr{\"o}dinger equation governed by the full time dependent
Hamiltonian (Eq. \ref{eq:H}):
\begin{equation}
 i\hbar \partial_t \, U_{t,t_0}=H(t)U_{t,t_0}\, , \qquad U_{t_0,t_0}=1 \, .
\end{equation}
At time $t_f$ the second measurement of $H_0$ is performed, and
the eigenvalue $e_m$ is obtained with probability:
\begin{equation}
 p(m|n) = \Tr \,P_m \rho_n(t_f) \, .
\end{equation}
Therefore the probability density to observe a certain value of
work $w_0$ is given by:
\begin{equation}
 p^0_{t_f,t_0}(w_0)= \sum_{m,n} \delta(w_0-[e_m-e_n])p(m|n)p_n \, .
\end{equation}
We use the superscript $0$ throughout this paper to indicate the
exclusive viewpoint. The same symbols, without the superscript $0$
denote the respective quantities within the inclusive viewpoint.

The Fourier transform of the probability density of work gives the
characteristic function of work
\begin{equation}
 G^0_{t_f,t_0}(u)= \int dw_0 p^0_{t_f,t_0}(w_0) e^{iuw_0}
\end{equation}
which allows quick derivations of fluctuation theorems and
nonequilibrium equalities. Performing calculations analogous to
those reported by \citet{THM_PRE08} we find the characteristic
function of work, in the form of a two point {\it quantum
correlation function}:
\begin{equation}
 G^0_{t_f,t_0}(u)= \Tr \, e^{i u H_0^H(t_f)} e^{-i u H_0}\bar
 \rho(t_0) \equiv \langle e^{i u H_0^H(t_f)} e^{-i u H_0}\rangle
\label{eq:G0}
\end{equation}
where $\bar \rho(t_0)$ is defined as:
\begin{equation}
\bar \rho(t_0) = \sum_n p_n \rho_n =\sum_n P_n \rho(t_0) P_n
\label{eq:bar-rho}
\end{equation}
and the superscript $H$ stands for Heisenberg representation,
i.e.:
\begin{equation}
H^H_0(t_f) = U^{\dagger}_{t_f,t_0} H_0 U_{t_f,t_0} \label{eq:H^H}
\end{equation}
This exclusive-work characteristic function $G^0_{t_f,t_0}$ should
be compared to the in\-clusive-work characteristic function
$G_{t_f,t_0}$ that is obtained when looking at the difference $w$ of
the outcomes $E_n(t_0)$ and $E_m(t_f)$ of measurements of the {\it
total} time dependent Hamiltonian $H(t)$. In this case one obtains
\citep{Lutz_PRE07,THM_PRE08}:
\begin{equation}
 G_{t_f,t_0}(u)= \Tr \, e^{i u H^H(t_f)} e^{-i u H_0}\bar \rho(t_0)
 \equiv \langle e^{i u H^H(t_f)} e^{-i u H_0}\rangle
\end{equation}
The difference lies in the distinct fact that $H_0^H(t_f)$ appears
in the exclusive approach in place of the full $H^H(t_f)$.

%%%%%%%%%%%%%%%%%%%%%%%%%%%%%%%%%%%%%%%%%%%%%%%%%%%%%%%%%%%%%%%%%%%%%%%%
\subsection{\label{sec:rev}Reversed protocol}
%%%%%%%%%%%%%%%%%%%%%%%%%%%%%%%%%%%%%%%%%%%%%%%%%%%%%%%%%%%%%%%%%%%%%%%%
Consider next the reversed protocol
\begin{equation}
\widetilde X(t) = X(t_f+t_0-t)
\label{eq:Xtilde}
\end{equation}
which consecutively assumes values as if time was reversed.
Let $\widetilde H(t)$ be the resulting Hamiltonian:
\begin{equation}
\widetilde H(t) = H_0 -  \widetilde X(t) Q
\label{eq:Htilde}
\end{equation}
The characteristic function of work now reads:
\begin{equation}
\widetilde G_{t_f,t_0}(u)= \Tr \, e^{i u \widetilde H^H(t_f)}
e^{-i u H_0}\bar \rho(t_0) \equiv \langle e^{i u\widetilde
H_0^H(t_f)} e^{-i u H_0}\rangle
\end{equation}
where
\begin{equation}
\widetilde H^H_0(t_f) = \widetilde U^{\dagger}_{t_f,t_0} H_0
\widetilde U_{t_f,t_0} \label{eq:H^H-tilde}
\end{equation}
and $\widetilde U_{t_f,t_0}$ is the time evolution operator
generated by $\widetilde H(t)$:
\begin{equation}
 i\hbar \partial_t \,\widetilde U_{t,t_0}=\widetilde H(t)\widetilde
 U_{t,t_0} \qquad \widetilde U_{t_0,t_0}=1
\end{equation}
Assuming that the Hamiltonian $H(t)$ is invariant under time
reversal i.e.,\footnote{Here we assume that the Hamiltonian does
not depend on any odd parameter, e.g., a magnetic field. Treating
that case is straightforward and amounts to reverse the sign of
the odd parameter in the r.h.s. of Eq. (\ref{eq:Theta-H-Theta}),
see \citep{Andrieux_NJP09}.}
\begin{equation}
 \Theta H(t) \Theta^{-1} = H(t)\, ,
\label{eq:Theta-H-Theta}
\end{equation}
where $\Theta$ is the antiunitary time reversal operator
\citep{MessiahBook},
the time evolution operators associated to the forward and
backward protocols are related by the following important relation,
see \ref{app:theta}:
\begin{equation}
U_{t_0,t_f}= U_{t_f,t_0}^{\dagger} = \Theta \widetilde U_{t_f,t_0} \Theta^{-1}
\, .
\label{eq:Theta-U-Theta}
\end{equation}
In the following section we will derive the quantum version of Eq.
(\ref{eq:BKE}) and its associated work fluctuation theorem. This
will be accomplished by choosing the initial density matrix to be
a Gibbs canonical state. In Sec. \ref{sec:mu-can} we will,
instead, assume an initial microcanonical state.

%%%%%%%%%%%%%%%%%%%%%%%%%%%%%%%%%%%%%%%%%%%%%%%%%%%%%%%%%%%%%%%%%%%%%%%%
\section{\label{sec:can}Canonical initial state}
%%%%%%%%%%%%%%%%%%%%%%%%%%%%%%%%%%%%%%%%%%%%%%%%%%%%%%%%%%%%%%%%%%%%%%%%
For a system staying at time $t_0$ in a canonical Gibbs state:
\begin{equation}
 \rho(t_0)=\bar \rho(t_0)= e^{-\beta H_0}/Z_0
\label{eq:rho-can}
\end{equation}
where $Z_0= \Tr\, e^{-\beta H_0}$, $\bar \rho (t_0)$ coincides
with $\rho(t_0)$ because the latter is diagonal with respect to the
eigenbasis of $H_0$ (see Eq. \ref{eq:bar-rho}). Plugging Eq.
(\ref{eq:rho-can}) into (\ref{eq:G0}), we obtain:
\begin{equation}
 G^0_{t_f,t_0}(\beta;u) = \Tr \, e^{i u H_0^H(t_f)} e^{-i u H_0}
 e^{-\beta H_0}/Z_0
\label{eq:G0-can}
\end{equation}
where for completeness we have listed the dependence upon $\beta$
among the arguments of $ G^0_{t_f,t_0}$. The quantum version of Eq.
(\ref{eq:BKE}) immediately follows by setting $u=i\beta$:
\begin{align}
\langle e^{-\beta w_0}\rangle =G^0_{t_f,t_0}(\beta;i\beta)= \Tr \,
e^{-\beta H_0^H(t_f)}/Z_0=\Tr \, e^{-\beta H_0}/Z_0=1
\label{eq:BKE-quantum}
\end{align}
where in the third equation we have used Eq. (\ref{eq:H^H}), the
cyclic property of the trace and the unitarity of the time
evolution operator: $U^{\dagger}_{t_f,t_0} U_{t_f,t_0}=1$.

Moreover we find the following important relation between $G^0_{t_f,t_0}$
and $\widetilde G^0_{t_f,t_0}$, see \ref{app:G}:
\begin{equation}
 G^0_{t_f,t_0}(\beta;u)= \widetilde G^0_{t_f,t_0}(\beta;-u+i\beta) \, .
\label{eq:G=Gtilde}
\end{equation}

By means of inverse Fourier transform, the following quantum
Bochkov-Kuzovlev fluctuation relation between the forward and
backward work probability density functions is obtained:
 \begin{equation}
\frac{p^0_{t_f,t_0}(\beta;w_0)}{\widetilde
p^0_{t_f,t_0}(\beta;-w_0)} = e^{\beta w_0} \, .
\label{eq:BK-FT}
\end{equation}
This must be compared to the quantum Tasaki-Crooks relation that
is obtained within the inclusive viewpoint \citep{TH_JPA07}:
 \begin{equation}
\frac{p_{t_f,t_0}(\beta;w)}{\widetilde p_{t_f,t_0}(\beta;-w)} =
e^{\beta (w-\Delta F)} \label{eq:TC-FT}
\end{equation}
where, in contrast to Eq. (\ref{eq:BK-FT}) the term $\Delta
F=-\beta^{-1}[ \ln {\Tr \, e^{-\beta H(t)}}- \ln {\Tr \, e^{-\beta
H_0}}]$, appears.

%%%%%%%%%%%%%%%%%%%%%%%
\subsection{Remarks}
Eqs. (\ref{eq:BKE-quantum}, \ref{eq:BK-FT}) constitute original
quantum results that do not appear in the works of
\citet{BochkovKuzovlev77,BochkovKuzovlev81}. In the {\it classical}
case they found a fluctuation theorem similar to Eq.
(\ref{eq:BK-FT}), reading:
 \begin{equation}
\frac{P[Q(\tau);X(\tau)]}{P[\varepsilon \widetilde Q(\tau) ;\varepsilon
\widetilde X(\tau)]} = \exp\left[\beta \int_{t_0}^{t_f} X(\tau)\dot
Q(\tau)\right]
\label{eq:BK-FT-QX}
\end{equation}
where $P[Q(\tau);X(\tau)]$ is the probability density {\it
functional} to observe a certain trajectory $Q(\tau)$ given a
certain protocol $X(\tau)$. Here $Q(\tau)$ is a short hand notation
for
$Q(\mathbf{q}(\mathbf{q}_0,\mathbf{p}_0,\tau),\mathbf{p}(\mathbf{q}_0,\mathbf{p}
_0,\tau))$, see Eq. (\ref{eq:H-cl}), where $(\mathbf{q}
(\mathbf{q}_0,\mathbf{p}_0,\tau),\mathbf{p}(\mathbf{q}_0,\mathbf{p}_0,\tau))$
is the evolved initial condition $\mathbf{q}_0,\mathbf{p}_0$ at some
time $\tau \in [t_0,t_f]$, for a certain protocol $X(\tau)$. The
symbol $\varepsilon$ denotes the parity of the observable $Q$ under
time reversal (assumed to be equal to $1$ in this paper). The symbol
$\sim$ denotes quantities referring to the reversed protocol. The
{\em classical} probability of work $p^{cl,0}_{t_f,t_0}(W_0)$ is
obtained from the $Q$-trajectory probability density functional
$P[Q(\tau);X(\tau)]$ via the formula:
 \begin{equation}
p^{cl,0}_{t_f,t_0}(W_0)=
\int \mathcal{D}Q(\tau) \,P[Q(\tau);X(\tau)] \, 
\delta\left[{W_0-\int_{t_0}^{t_f} X(\tau)\dot Q(\tau)} \right]
\end{equation}
where the integration is a functional integration over all possible trajectories
such that $\int_{t_0}^{t_f} X(\tau)\dot Q(\tau)=W_0$. With this formula one
finds from Eq. (\ref{eq:BK-FT-QX}) the exclusive version of the classical Crooks
fluctuation theorem for the work probability densities \citep{Jarz_JStatMech07}
\begin{equation}
p^{cl,0}_{t_f,t_0}(\beta;W_0)=\widetilde p^{cl,0}_{t_f,t_0}(\beta;-W_0)e^{\beta
W_0} \,.
\end{equation}
Notably, a quantum version of Eq. (\ref{eq:BK-FT-QX}) does not
exists because:... ``in the quantum case it is impossible to
introduce unambiguously a [...] probability functional''
\citep{BochkovKuzovlev81}. It is only by giving up the idea of true
quantum trajectories and embracing instead the two-point measurement
approach that the quantum exclusive fluctuation theorem Eq.
(\ref{eq:BK-FT}) can be obtained, and has been obtained here, for
the first time.

%%%%%%%%%%%%%%%%%%%%%%%%%%%%%%%%%%%%%%%%%%%%%%%%%%%%%%%%%%%%%%%%%%%%%%%%
\section{\label{sec:mu-can}Microcanonical initial state}
%%%%%%%%%%%%%%%%%%%%%%%%%%%%%%%%%%%%%%%%%%%%%%%%%%%%%%%%%%%%%%%%%%%%%%%%
We consider next an initial microcanonical initial  state of energy
$E$, that can formally be expressed as:
\begin{equation}
 \rho(t_0)=\bar \rho(t_0)= \delta(H_0-E)/\Omega_0(E) \, ,
\label{eq:rho-mu-can}
\end{equation}
% DEFINE HERE OMEGA_0 !
wherein $\Omega_0(E)= \Tr \, \delta(H_0-E)$.  Actually one has to
replace the singular Dirac function $\delta(x)$ by a smooth function
sharply peaked around $x=0$, but with infinite support. A normalized
gaussian with arbitrarily small width serves this purpose well.

With this choice of initial condition, the characteristic function
of work reads:
 \begin{align}
 G^0_{t_f,t_0}(E;u) &= \Tr \, e^{i u H_0^H(t_f)} e^{-i u H_0} \delta
 (H_0-E)/\Omega_0(E) \nonumber \\
&= \Tr \, e^{i u [H_0^H(t_f)-E]} \delta (H_0-E)/\Omega_0(E)
\label{eq:G0-mu-can}
\end{align}
where for completeness we listed the dependence upon $E$ among the
arguments of $ G^0_{t_f,t_0}$. By applying the inverse Fourier
transform we obtain:
\begin{equation}
  p^0_{t_f,t_0}(E;w_0)= \Tr \, \delta(H_0^H(t_f)-E-w_0)\delta
  (H_0-E)/\Omega_0(E)
\label{eq:p-mu-can}
\end{equation}
Likewise, for the reversed protocol,
 \begin{equation}
\widetilde  p^0_{t_f,t_0}(E;w_0)= \Tr \, \delta(\widetilde
H_0^H(t_f)-E-w_0)\delta (H_0-E)/\Omega_0(E)
\label{eq:p-mu-can-Rev}
\end{equation}
is found.

We then find the following relation between the forward and
backward work probability densities, see \ref{app:OmegaP}:
\begin{equation}
\Omega_0(E) p^0_{t_f,t_0}(E;w_0) = \Omega_0(E+w_0) \widetilde
p^0_{t_f,t_0}(E+w_0;-w_0)
\label{eq:OmegaP=OmegaPtilde}
\end{equation}

Then, the quantum microcanonical fluctuation theorem reads,
within the exclusive viewpoint:
 \begin{equation}
\frac{p^0_{t_f,t_0}(E;w_0)}{\widetilde p^0_{t_f,t_0}(E+w_0;-w_0)}
= \frac{\Omega_0(E+w_0)}{\Omega_0(E)}\label{eq:BK-FT-mu}
\end{equation}
This must be compared to the quantum microcanonical fluctuation
theorem, obtained within the inclusive viewpoint \citep{THM_PRE08}
 \begin{equation}
\frac{p_{t_f,t_0}(E;w)}{\widetilde p_{t_f,t_0}(E+w;-w)} =
\frac{\Omega_f(E+w)}{\Omega_0(E)}\label{eq:TC-FT-mu}
\end{equation}
The difference lies in the fact that within the exclusive
viewpoint the densities of states at the final energy $E+w_0$, is
determined by the unperturbed Hamiltonian, i.e.,  $\Omega_0(E+w_0)=\Tr
\, \delta (H_0-(E+w_0))$, whereas it results from the total
Hamiltonian in the inclusive approach: $\Omega_f(E+w)=\Tr \,
\delta (H(t_f)-E-w)$.

Eq. (\ref{eq:TC-FT-mu}) was first obtained
within the classical framework by \citep{Cleuren_PRL06}. It is not
difficult to see that Eq. (\ref{eq:BK-FT-mu}) holds classically as
well.

%%%%%%%%%%%%%%%%%%%%%%%
\subsection{Remarks}

Just as  Eq. (\ref{eq:BK-FT}), this Eq. (\ref{eq:BK-FT-mu}) is a new
result that was not reported before by
\citet{BochkovKuzovlev77,BochkovKuzovlev81}. It is very interesting
to notice, however, that those authors already put forward a {\it
classical} fluctuation theorem for the microcanonical ensemble,
which can be recast in the form \citep{BochkovKuzovlev81}:
\begin{equation}
 \frac{P[I(\tau);X(\tau);E]}{P[-\varepsilon \widetilde I(\tau);\varepsilon
\widetilde X(\tau);E+W_0]} =\frac{\Omega_0(E+W_0)}{\Omega_0(E)}
\label{eq:BK-FT-mu-IX}
\end{equation}
where $P[I(\tau);X(\tau);E]$ is the probability density {\it functional} to
observe a certain trajectory $I(\tau)$ given a certain protocol and an initial
microcanonical ensemble of energy $E$. Here
\begin{equation}
I(\tau) = \dot
Q(\mathbf{q}(\mathbf{q}_0,\mathbf{p}_0,\tau),\mathbf{p}(\mathbf{q}_0,\mathbf{p}
_0,\tau))
\end{equation}
denotes the {\it current}.
By functional integration the classical microcanonical theorem for the pdf of
work
\begin{equation}
\frac{p^{cl,0}_{t_f,t_0}(E,W_0)}{\widetilde p^{cl,0}_{t_f,t_0}(E+W_0,W_0)}=
\frac{\Omega_0(E+W_0)}{\Omega_0(E)}
\end{equation}
is obtained from (\ref{eq:BK-FT-mu-IX}) in the same way as
(\ref{eq:BK-FT}) follows from (\ref{eq:BK-FT-QX}). However the
quantum version of (\ref{eq:BK-FT-mu-IX}) does not exists and the
derivation of the quantum microcanonical fluctuation theorem
(\ref{eq:BK-FT-mu}) is indeed only possible if the two-point
measurement approach is adopted.

The fluctuation relations of Eqs. (\ref{eq:BK-FT-mu},
\ref{eq:TC-FT-mu}) can be further expressed in terms of entropy,
according to the rules of statistical mechanics.
 Following \cite{Gibbs02} two different prescriptions are found in
 textbooks to obtain the entropy associated to the microcanonical
 ensemble:
\begin{align}
 s(E)&= k_B \ln \Omega(E) = \Tr\,\delta(H-E) \\
 S(E)&= k_B \ln \Phi(E) = \Tr\, \theta(H-E)
\end{align}
The two definitions coincide for large systems with short range
interactions among their constituents, but may substantially differ
if the size of the system under study is small. It is by now clear
that, of the two, only the second -- customarily called ``Hertz
entropy" -- is the fundamentally correct one
(\cite{Hertz1,Hertz2,Schluter48,Pearson85,Campisi05,Campisi08,Campisi_AJP10,
Dunkel06}). \footnote{It is interesting to notice that Einstein was
well aware of
  the works of \citet{Hertz1,Hertz2} which he praised as excellent
  (``vortrefflich'') \citep{Einstein11}.}
Using the microcanonical expression for the
temperature $k_B T(E)=(\partial S(E)/ \partial E)^{-1}=\Phi(E)/\Omega(E)$, we can re-express the
quantum microcanonical Bochkov-Kuzovlev fluctuation relation in
terms of entropy and temperature as:
 \begin{equation}
\frac{p^0_{t_f,t_0}(E;w_0)}{\widetilde p^0_{t_f,t_0}(E+w_0;-w_0)}
= \frac{T_0(E)}{T_0(E+w_0)}e^{[S_0(E+w_0)-S_0(E)]/k_B} \label{eq:BK-FTb}
\end{equation}
where the subscript $0$ in $T$ and $S$ denotes that these
quantities are calculated for the unperturbed Hamiltonian $H_0$.
Likewise, adopting the inclusive viewpoint one obtains:
 \begin{equation}
\frac{p_{t_f,t_0}(E;w)}{\widetilde p_{t_f,t_0}(E+w;-w)} =
\frac{T_0(E)}{T_f(E+w)}e^{[S_f(E+w)-S_0(E)]/k_B} \label{eq:TC-FTb}
\end{equation}
where the subscript $f$ in $T$ and $S$ denotes that these
quantities are calculated for the total final Hamiltonian
$H(t_f)$. \footnote{If instead of the microcanonical ensemble 
(\ref{eq:rho-mu-can}), the modified microcanonical ensemble $\varrho(t_0)=\theta(E-H_0)/[\Tr\, \theta(E-H_0)]$, \citep{Ruelle}
would be used as the initial equilibrium state, then the fluctuation
 theorem assumes the same form as in Eq. (\ref{eq:TC-FTb}), 
but without the ratio of temperatures \citep{THM_PRE08}.}

%%%%%%%%%%%%%%%%%%%%%%%%%%%%%%%%%%%%%%%%%%%%%%%%%%%%%%%%%%%%%%%%%%%%%%%%
\section{\label{sec:discussion}Discussion}
%%%%%%%%%%%%%%%%%%%%%%%%%%%%%%%%%%%%%%%%%%%%%%%%%%%%%%%%%%%%%%%%%%%%%%%%
We derived the quantum Bochkov-Kuzovlev identity as well as the
quantum canonical and microcanonical work fluctuation theorems within
the exclusive approach, and have elucidated their relations to the original
works of \citet{BochkovKuzovlev77,BochkovKuzovlev81}.
The extension of the corresponding classical theorems to the quantum
regime is only possible thanks to the proper definition of work as a
two-time quantum observable.
We close with two comments: \footnote{Similar remarks were made also
  within the classical framework by \cite{Jarz_CRPhys07}.}

1. For a cyclic process, $X(t_f)=X(t_0)$, inclusive and exclusive
work fluctuation theorems coincide. However in no way is it true
that the exclusive approach of Bochkov \&  Kuzovlev, adopted here,
is restricted to cyclic processes, as some authors have suggested
\citep{Allahverdyan_PRE05,Cohen_2005,Andrieux_PRL08}. As stressed in
the introduction, the difference of the two approaches originates
from  the different definitions of work, and is not related to
whether the process under study is cyclic or is not cyclic.

2. Within the inclusive approach it is natural to define the
\emph{dissipated work} as $w_{dis}=w-\Delta F$
\citep{Kawai_PRL07,Jarzynski_EPL09}. Then, the Jarzynski equality
(\ref{eq:JE}) can be rewritten as $\langle e^{-\beta w_{dis}}\rangle
=1$. This might make one believe that the exclusive work $w_0$
coincides with the dissipated work $w_{dis}$. This, though, would be
generally wrong. The dissipated work $w_{dis}$ is a stochastic
quantity whose statistics, given by
$p^{dis}_{t_f,t_0}(w_{dis})=p_{t_f,t_0}(w_{dis} +\Delta F)$, in
general does not coincide with the statistics of exclusive work
$w_0$, given by $p^{0}_{t_f,t_0}(w_0)$. See \ref{app:WdissPDF} for a
counterexample. Only for a cyclic process, for which $\Delta F=0$,
does the dissipated-work $w_{dis}$ coincide with the inclusive-work
$w$, which in turn coincides with the exclusive-work $w_0$.

%%%%%%%%%%%%%%%%%%%%%%%%%%%%%%%%%%%%%%%%%%%%%%%%%%%%%%%%%%%%%%%%%%%%%%%%%%
\subsection*{Acknowledgments}
%%%%%%%%%%%%%%%%%%%%%%%%%%%%%%%%%%%%%%%%%%%%%%%%%%%%%%%%%%%%%%%%%%%%%%%%%%
The authors gratefully acknowledge financial support from the German
Excellence Initiative via the {\it Nanosystems Initiative Munich}
(NIM), the Volkswagen Foundation (project I/80424), and the DFG via
the collaborative research center SFB-486, Project A-10, via the DFG
project no. HA1517/26--2.

%%%%%%%%%%%%%%%%%%%%%%%%%%%%%%%%%%%%%%%%%%%%%%%%%%%%%%%%%%%%%%%%%%%%%%%%%%
\appendix{\label{app:theta}Derivation of Eq. (\ref{eq:Theta-U-Theta})}
%%%%%%%%%%%%%%%%%%%%%%%%%%%%%%%%%%%%%%%%%%%%%%%%%%%%%%%%%%%%%%%%%%%%%%%%%%
The time evolution operator $\widetilde U_{t,t_0}$ can be expressed as
a time ordered product \citep{SchleichBook}:
\begin{equation}
 \widetilde U_{t_f,t_0}= \lim_{N\rightarrow \infty} e^{-i \widetilde H
   (t_{N})\tau} e^{-i \widetilde H (t_{N-2})\tau} \dots e^{-i
   \widetilde H (t_{1})\tau}
\end{equation}
where $\tau = (t_f-t_0)/N$, and $t_\nu=t_0 + \nu \tau$, for $\nu = 0,
\dots, N $ (note that $t_N=t_f$).
Due to Eqs. (\ref{eq:Xtilde}, \ref{eq:Htilde}), it is $\widetilde
H(t)=H(t_f+t_0-t)$, then:
\begin{equation}
 \widetilde U_{t_f,t_0}= \lim_{N\rightarrow \infty} e^{-i H
   (t_{1})\tau} e^{-i H(t_{2})\tau} \dots e^{-i H (t_{N})\tau}
\end{equation}
Therefore:
\begin{equation}
\Theta \widetilde U_{t_f,t_0} \Theta^{-1}= \lim_{N\rightarrow \infty}
\Theta e^{-i H (t_{1})\tau} \Theta^{-1}\Theta  e^{-i H(t_{2})\tau}
\Theta^{-1} \dots\Theta e^{-i H (t_{N})\tau} \Theta^{-1}
\end{equation}
where we inserted $\Theta^{-1}\Theta={\mathbb 1}$, $N-1$ times. Due to
the property (\ref{eq:Theta-H-Theta}) and the antilinearity of
$\Theta$ it is
\begin{equation}
 \Theta e^{-i H(t)\tau}\Theta^{-1}=e^{i H(t)\tau} \, .
\label{eq:Theta-e^iH-Theta}
\end{equation}
Using this equation, we find:
\begin{align}
\Theta \widetilde U_{t_f,t_0} \Theta^{-1} &= \lim_{N\rightarrow
  \infty}  e^{i H (t_{1})\tau} e^{i H(t_{2})\tau}  \dots e^{i H
  (t_{N})\tau} \\
&=  \lim_{N\rightarrow \infty}\left[ e^{-i H (t_{N})\tau} \dots e^{-i
    H(t_{2})\tau} e^{-i H (t_{1})\tau}\right]^{\dagger} \\
&= U_{t_f,t_0}^{\dagger} = U_{t_0,t_f} \, .
\label{eq:Theta-Utilde-Theta1}
\end{align}
In a similar way we also obtain:
\begin{equation}
U_{t_f,t_0}= \Theta \widetilde U_{t_f,t_0}^{\dagger} \Theta^{-1} \, .
\label{eq:Theta-Utilde-Theta2}
\end{equation}

%%%%%%%%%%%%%%%%%%%%%%%%%%%%%%%%%%%%%%%%%%%%%%%%%%%%%%%%%%%%%%%%%%%%%%%%%%
\appendix{\label{app:G}Derivation of Eq. (\ref{eq:G=Gtilde})}
%%%%%%%%%%%%%%%%%%%%%%%%%%%%%%%%%%%%%%%%%%%%%%%%%%%%%%%%%%%%%%%%%%%%%%%%%%
The exclusive-work characteristic function reads (\ref{eq:G0})
\begin{equation}
  G^0_{t_f,t_0}(\beta;u)=\Tr \,U_{t_f,t_0} e^{i u H_0}
  U^{\dagger}_{t_f,t_0} e^{-i u H_0} e^{-\beta H_0}/Z_0  \, .
\end{equation}
Using Eqs. (\ref{eq:Theta-Utilde-Theta1},\ref{eq:Theta-Utilde-Theta2}), we
obtain:
\begin{equation}
  G^0_{t_f,t_0}(\beta;u)=\Tr \,\Theta \widetilde U_{t_f,t_0}^\dagger
  \Theta^{-1} e^{i u H_0} \Theta \widetilde U_{t_f,t_0} \Theta
  ^{-1}e^{-i u H_0} e^{-\beta H_0} \Theta \Theta ^{-1}/Z_0 \nonumber
\end{equation}
where we have inserted $\Theta\Theta^{-1}={\mathbb 1}$ at the right
end. By multiplying Eq. (\ref{eq:Theta-e^iH-Theta}) by $\Theta^{-1}$
to the left and by $\Theta$ to the right, we have (replacing $\tau$
with $u$)
\begin{equation}
 \Theta^{-1} e^{i H(t)u} \Theta=e^{-i H(t)u} \, ,
\end{equation}
therefore:
\begin{align}
  G^0_{t_f,t_0}(\beta;u)&=\Tr \,\Theta \widetilde U_{t_f,t_0}^\dagger
  e^{-i u H_0} \widetilde U_{t_f,t_0} e^{i u H_0} e^{-\beta H_0}
  \Theta^{-1} /Z_0  \, .
\end{align}
The antilinearity of $\Theta$ implies, for any trace class operator $A$:
\begin{equation}
\Tr \,\Theta A \Theta^{-1} = \Tr \, A^{\dagger} \, .
\label{eq:TrThetaATheta}
\end{equation}
Therefore:
\begin{equation}
  G^0_{t_f,t_0}(\beta;u)=\Tr \,e^{-\beta H_0}  e^{-i u H_0}\widetilde
  U_{t_f,t_0}^\dagger e^{i u H_0}  \widetilde U_{t_f,t_0}  /Z_0
\end{equation}
Using the cyclic property of the trace finally leads to
\begin{align}
  G^0_{t_f,t_0}(\beta;u)&=\Tr \, \widetilde U_{t_f,t_0}e^{i(-u+i\beta)
    H_0}  \widetilde U_{t_f,t_0}^\dagger e^{-i(-u+i\beta) H_0}
  e^{-\beta H_0}  /Z_0\\
&=\widetilde G^0_{t_f,t_0}(\beta;-u+i\beta)\, .
\end{align}

%%%%%%%%%%%%%%%%%%%%%%%%%%%%%%%%%%%%%%%%%%%%%%%%%%%%%%%%%%%%%%%%%%%%%%%%%%
\appendix{\label{app:OmegaP}Derivation of
  Eq. (\ref{eq:OmegaP=OmegaPtilde})}
%%%%%%%%%%%%%%%%%%%%%%%%%%%%%%%%%%%%%%%%%%%%%%%%%%%%%%%%%%%%%%%%%%%%%%%%%%
The microcanonical exclusive-work probability density function reads,
Eq. (\ref{eq:p-mu-can}):
\begin{align}
p^0_{t_f,t_0}(E;w_0)&= \Tr \,
\delta(H_0^H(t_f)-E-w_0)\delta(H_0-E)/\Omega_0(E)\\
&= \Tr \, U_{t_f,t_0}^{\dagger}\delta(H_0-E-w_0) U_{t_f,t_0}\delta
(H_0-E)/\Omega_0(E)
\end{align}
Employing Eqs. (\ref{eq:Theta-Utilde-Theta1},\ref{eq:Theta-Utilde-Theta2}), then
leads to:
\begin{align}
\Omega_0(E) p^0_{t_f,t_0}(E;w_0)&=  \Tr \,\Theta \widetilde
U_{t_f,t_0} \Theta^{-1}\delta(H_0-E-w_0) \Theta \widetilde
U_{t_f,t_0}^{\dagger} \Theta^{-1} \delta (H_0-E)  \Theta \Theta^{-1}
\end{align}
where we have inserted $\Theta \Theta^{-1}=\mathbb{1}$ at the end.
Being the Dirac delta a {\it real} function we have
\begin{equation}
\Theta^{-1} \delta (H_0-E) \Theta= \delta (H_0-E)
\end{equation}
because $H_0$ is assumed to be invariant under time reversal. Then:
\begin{equation}
\Omega_0(E) p^0_{t_f,t_0}(E;w_0)=  \Tr \, \Theta \widetilde
U_{t_f,t_0} \delta(H_0-E-w_0)\widetilde U_{t_f,t_0}^{\dagger} \delta
(H_0-E) \Theta^{-1} \, .
\end{equation}
Using Eq. (\ref{eq:TrThetaATheta}), we obtain:
\begin{align}
\Omega_0(E) p^0_{t_f,t_0}(E;w_0)
&= \Tr \,\delta (H_0-E)\widetilde U_{t_f,t_0}\delta(H_0-E-w_0)
\widetilde U_{t_f,t_0}^{\dagger} \, .
\end{align}
Thanks to the cyclic property of the trace one finally arrives at:
\begin{align}
\Omega_0(E) p^0_{t_f,t_0}(E;w_0)
&= \Tr \,\widetilde U_{t_f,t_0}^{\dagger} \delta (H_0-E)\widetilde
U_{t_f,t_0}\delta(H_0-E-w_0) \\
&= \Omega_0(E+w_0) \widetilde p^0_{t_f,t_0}(E+w_0;w_0)  \, .
\end{align}

%%%%%%%%%%%%%%%%%%%%%%%%%%%%%%%%%%%%%%%%%%%%%%%%%%%%%%%%%%%%%%%%%%%%%%%%%%
\appendix{\label{app:WdissPDF}Comparison between dissipated-work and
  exclusive-work pdf's}
%%%%%%%%%%%%%%%%%%%%%%%%%%%%%%%%%%%%%%%%%%%%%%%%%%%%%%%%%%%%%%%%%%%%%%%%%%
In this appendix we provide an example that shows that the
dissipated-work $w_{dis}$ and the inclusive work $w_0$ are distinct
stochastic quantities with different statistical properties. To this
end we show that their probability density functions may have
different first and second moments.
We consider a driven quantum harmonic oscillator of unit mass and unit
angular frequency:
\begin{equation}
H(t) = p^2/2 + q^2/2 -X(t)q
\end{equation}
For simplicity we assume $t_0=0$, $X(t_0)=0$, and we chose units in such a way
that $\hbar=1$. Let $|n,t\rangle$ denote the instantaneous
eigenvectors of $H(t)$ corresponding to the instantaneous eigenvalues
$E_n(t)= (n+1/2)-X^2(t)/2$.

%%%%%%%%%%%%%%%%%%%%%%%%%%%%%%%%%%%%%%%%%%%%%%%%%%%%%%%%%%%%%%%%%%%%%%%%%%
\subsection{The probability density of dissipated-work}
%%%%%%%%%%%%%%%%%%%%%%%%%%%%%%%%%%%%%%%%%%%%%%%%%%%%%%%%%%%%%%%%%%%%%%%%%%
The probability density function (pdf) of inclusive work,
corresponding to an initial canonical state, is
\begin{equation}
p_{t,0}(w)=\sum_{m n} \delta(w- m + n +X^2(t)/2)|a_{mn}|^2 e^{-\beta
  (n+1/2)}/Z(0)
\label{eq:in-pdf}
\end{equation}
where $Z(0)=\sum_n e^{-\beta ( n+1/2)}$ is the initial partition
function, and $|a_{nm}|^2$ are the probabilities to make a transition
between two eigenstates of the total Hamiltonian
 \begin{equation}
|a_{mn}|^2=|\langle m,t|U_{t,0}|n,0\rangle|^2
\end{equation}
where we have set $t_0=0$ and $t_f=t$. According to
\citet{TBH_PRE08,TBH_erratum} the mean value and the variance of the
inclusive work pdf (\ref{eq:in-pdf}) are given by
\begin{align}
\langle w\rangle&=\int dx \, x \, p_{t,0}(x) =  L(t)  -X^2(t)/2\\
\langle \Delta w^2\rangle&=\int dx [x-L(t)]^2 p_{t,0}(x)=2  U L(t)
\end{align}
where $U= \sum_n (n+1/2)e^{-\beta ( n+1/2)}/Z_0$ is the initial
average energy, and \footnote{In \citep{TBH_PRE08,TBH_erratum} $L$ is
  given as $L(t)= |\int_0^t ds {\dot f}(s) e^{is}|^2$, where
  $f=-X/\sqrt{2}$. It is a matter of elementary calculus to check that
  this expression coincides with Eq. (\ref{eq:L}).}
\begin{equation}
L(t)= C(t)^2/2 +[S(t)-X(t)]^2/2
\label{eq:L}
\end{equation}
where
\begin{align}
S(t)=\int_0^t ds X(s) \sin(t-s)\, , \qquad C(t)=\int_0^t ds X(s)
\cos(t-s) \,.
\end{align}

The partition function of work at the final time $t$ is $Z(t)=Z(0)
e^{\beta X^2(t)/2}$, therefore the free energy difference $\Delta F =
-\beta^{-1} \ln Z(t)/Z_0$  is given by $\Delta F=-X^2(t)/2$
\citep{TBH_PRE08}.
Hence the dissipated work is
\begin{equation}
w_{dis}=w+X^2(t)/2 \,.
\end{equation}
Accordingly the dissipated work pdf is
\begin{align}
p^{dis}_{t,0}(w_{dis})&=  p_{t,0}(w)= p_{t,0}(w_{dis}-X^2(t)/2) \nonumber \\
&=\sum_{m n} \delta(w_{dis}- m + n)|a_{mn}|^2 e^{-\beta  (n+1/2)}/Z(0) \, .
\label{eq:dis-work-pdf}
\end{align}
It immediately follows that
\begin{align}
\langle w_{dis}\rangle&=\int dx \, x \, p^{dis}_{t,0}(x) =  L(t)\\
\langle \Delta w_{dis}^2\rangle&=\int dx [x-L(t)]^2 p^{dis}_{t,0}(x)=2
U L(t) \,.
\end{align}
Note that, as it should be, $\langle w_{dis}\rangle \geq 0$.

%%%%%%%%%%%%%%%%%%%%%%%%%%%%%%%%%%%%%%%%%%%%%%%%%%%%%%%%%%%%%%%%%%%%%%%%%%
\subsection{The probability density of exclusive-work}
%%%%%%%%%%%%%%%%%%%%%%%%%%%%%%%%%%%%%%%%%%%%%%%%%%%%%%%%%%%%%%%%%%%%%%%%%%
The exclusive-work pdf is given by
\begin{equation}
p^0_{t,0}(w_0)=\sum_{m n} \delta(w_0- m + n)|a^0_{mn}|^2 e^{-\beta
  (n+1/2)}/Z(0)
\label{eq:ex-work-pdf}
\end{equation}
where $|a^0_{mn}|^2$ denotes the probability to make a transition
between two states of the unperturbed Hamiltonian:
 \begin{equation}
|a^0_{mn}|^2=|\langle m,0|U_{t,0}|n,0\rangle|^2 \,.
\end{equation}
It is known \citep{Husimi53,Campisi_PRE08} that the transition
probabilities $|a_{mn}|^2$ depend on the time $t$ at which the second
measurement is performed, via the function $L(t)$, that is the
$|a_{mn}|^2$ are of the form $|a_{mn}|^2=f_{nm}[L(t)]$, for certain
functions $f_{nm}$ that need not be specified here.
Using Wigner functions to calculate the transition probabilities as in
\citep[Appendix]{Campisi_PRE08}, we notice that the transition
probabilities $|a^0_{mn}|^2$ are obtained from the same expression as
of $|a_{mn}|^2$, with the only difference that  $L(t)$ is replaced by
\begin{equation}
L_0(t)= C^2(t)/2 +S^2(t)/2
\end{equation}
that is $|a^0_{mn}|^2=f_{nm}[L_0(t)]$. Therefore the exclusive-work
pdf (\ref{eq:ex-work-pdf}) is obtained from the dissipated-work pdf
(\ref{eq:dis-work-pdf}) simply by replacing $L(t)$ with $L_0(t)$. It
follows immediately that
\begin{align}
\langle w_0 \rangle&=\int dx \, x \,p^0_{t,0}(x) =  L_0(t)\\
\langle \Delta w_0^2\rangle&=\int dx [x-L_0(t)]^2 p^0_{t,0}(x)=2  U L_0(t)
\end{align}
Note that, as expected, $\langle w_0\rangle\geq 0$.

For the specific protocol
\begin{equation}
X(t)= 2 \sin(t)
\label{eq:X}
\end{equation}
we find
\begin{align}
L(t)-L_0(t)=t\sin(2t)
\end{align}
which is apparently different from zero  except for integer multiples
of $\pi/2$. Thus for any duration $t$ of the protocol (\ref{eq:X})
that is not an integer multiple of $\pi/2$, $L_0\neq L$. Accordingly
the first and second moments of $p^{dis}_{t,0}$ and $p^0_{t,0}$
differ, meaning that $w_{dis}$ and $w_0$ are distinct stochastic
variables with different statistical properties.

It should be stressed that analogous results are found also for a
classical driven harmonic oscillator. The statistics of
dissipated-work and of exclusive-work generally {\it differ}, this
fact holds true both quantum-mechanically and classically.

%%%%%%%%%%%%%%%%%%%%%%%%%%%%%%%%%%%%%%%%%%%%%%%%%%%%%%%%%%%%%%%%%%%%%%%%%%
%%%%%%%%%%%%%%%%%%%%%%%%%%%%%%%%%%%%%%%%%%%%%%%%%%%%%%%%%%%%%%%%%%%%%%%%%%
\bibliographystyle{Harvard}

\end{document}